\newcommand{\beginsupplement}{%
        \setcounter{table}{0}
        \renewcommand{\thetable}{S\arabic{table}}%
        \setcounter{figure}{0}
        \renewcommand{\thefigure}{S\arabic{figure}}%
     }

\documentclass[runningheads]{llncs}
\usepackage{orcidlink}
\DeclareUnicodeCharacter{2212}{-}

\usepackage{graphicx,adjustbox,amsmath}
\usepackage{hyperref}
\usepackage{amsmath}
\usepackage{amssymb}
\usepackage{hyperref}

\begin{document}
\title{Estimation of 3T MR images from 1.5T images regularized with Physics based Constraint
\thanks{This work is financially supported by Ministry of Electronics and Information Technology, India}}
\titlerunning{Estimate HF images using physics based constraint}
%
\author{Prabhjot Kaur `I\inst{1}'\orcidlink{0000-0001-5416-0219} \and
Atul Singh Minhas\inst{2,3}\orcidlink{0000-0002-7922-3236} \and
Chirag Kamal Ahuja\inst{3}\orcidlink{0000-0003-0734-3252} \and Anil Kumar Sao\inst{4}\orcidlink{0000-0001-5316-5528} }
%
\authorrunning{P. Kaur et al.}
%
\institute{Indian Institute of Technology Mandi, India  \and
Magnetica, Australia, 
\and
Postgraduate Institute of Medical Education and Research, Chandigarh, India  \and
Indian Institute of Technology Bhilai, India\\\email{kaurprabhjotinresearch@gmail.com }
}

%
\maketitle              
\vspace{-1cm}

\begin{abstract}
Limited accessibility to high field MRI scanners (such as 7T, 11T) has motivated the development of post-processing methods to improve low field images. Several existing post-processing methods have shown the feasibility to improve 3T images to produce 7T-\textit{like} images~\cite{dlmia2016CNN3t7t,miccai2018Dual3t7t}. It has been observed that improving lower field (LF, $\leq1.5$T) images comes with additional challenges due to poor image quality such as the function mapping 1.5T and higher field (HF, 3T) images is more complex than the function relating 3T and 7T images~\cite{harrymiccai}. Except for~\cite{harrymiccai}, no method has been addressed to improve $\leq$1.5T MRI images. Further, most of the existing methods~\cite{dlmia2016CNN3t7t,miccai2018Dual3t7t} including~\cite{harrymiccai} require example images, and also often rely on pixel to pixel correspondences between LF and HF images which are usually inaccurate for $\leq$1.5T images. 
The focus of this paper is to address the unsupervised framework for quality improvement of 1.5T images and avoid the expensive requirements of example images and associated image registration. The LF and HF images are assumed to be related by a linear transformation (LT).
The unknown HF image and unknown LT are estimated in alternate minimization framework. Further, a physics based constraint is proposed that provides an additional non-linear function relating LF and HF images in order to achieve the desired high contrast in estimated HF image. This constraint exploits the fact that the T1 relaxation time of tissues increases with increase in field strength, and if it is incorporated in the LF acquisition the HF contrast can be simulated. The experimental results  demonstrate that the proposed approach provides processed 1.5T images, i.e., estimated 3T-\textit{like} images with improved image quality, 
and is comparably better than the existing methods addressing similar problems. The improvement in image quality is also shown to provide better tissue segmentation and volume quantification as compared to scanner acquired 1.5T images. The same set of experiments have also been conducted for 0.25T images to estimate 1.5T images, and demonstrate the advantages of proposed work.

\end{abstract}
\section{Introduction}
Magnetic resonance imaging (MRI) has seen a tremendous growth over the past three decades as a preferred diagnostic imaging modality. Starting with the 0.25T MRI scanners in 1980s~\cite{LowFieldMriSignalPhysicsPerspective}, clinical scanners have emerged from 1.5T and 3T to the recently approved 7T scanners~\cite{7TisBetter}. Primary reason for the preference of higher magnetic field strength (FS) is the better image quality but it comes with high cost. Generally, the cost of clinical MRI scanners increases at USD 1million/Tesla~\cite{howlowcanwego}. Therefore, there is a clear divide in the distribution of  MRI scanners across different countries in the world. With the FDA approval of clinical use of 7T MRI scanners- 3T and 7T MRI scanners have become preferred scanners in developed countries but inexpensive low field MRI scanners ($<$1T) and mid field MRI scanners (1.5T) are still a popular choice in developing countries~\cite{LowFieldMriSignalPhysicsPerspective,harrymiccai}.

Since the past decade, there is a growing interest towards improving the quality of images acquired with low FS MRI scanners by learning the features responsible for image quality from high FS MRI scanners~\cite{7TisBetter,7TInClinics,ICIP18}. 
Earlier methods in such image translational problems synthesized target image contrast using paired example images~\cite{mimecs}. The first work to address estimation of 7T-\textit{like} images was reported in 2016~\cite{miccai2016RFSR3t7t}. It exploits canonical correlation analysis (CCA)~\cite{cca} to relate paired 3T-7T images in a more correlated space. With advent of deep learning, 3D convolutional neural network (CNN) was proposed to learn non-linear mapping between 3T and 7T images~\cite{dlmia2016CNN3t7t}. Different architectures of deep learning  were addressed with better performances than previous methods~\cite{miccai2018Dual3t7t,MICCAI2017,ICIP18,miccaiwe19}. 
However, except Lin et al~\cite{harrymiccai}, none of the existing approaches address the problem of improving the image quality of $\leq$1.5T images to estimate high quality 3T-like MR images. Notably, to the best of our knowledge no approach is reported in literature to estimate the 3T (or 1.5T-like) images from 1.5T (or 0.25T) images. This problem is particularly challenging because of the severity of degradation present in $\leq$1T MR images~\cite{harrymiccai}. In \cite{harrymiccai}, the mapping between example 3T and simulated 0.36T images using CNNs to estimate 3T from scanner-acquired 0.36T images is learned, and it requires the a-priori knowledge of distribution of tissue-specific SNR for given FS. 
Since the databases for example images are scarce and SNR of tissues is not known a-priori~\cite{goldentableT1}, the unsupervised methods needs to be explored to improve $\leq$1.5T images.


In this work, we address the problem of estimating 3T-like images ($\mathbf{x}$) from 1.5T images ($\mathbf{y}$) in an unsupervised manner. To our best knowledge, this is the first work to develop a method for improving $\leq$1.5T images to estimate HF images without requiring example images. The proposed method formulates the estimation of 3T HF images as an inverse problem: $\mathbf{y}=f(\mathbf{x})$, where $f(.)$ is the unknown degradation model. The novel contributions of our work are as follows: (i) the alternate minimization (AM) framework is formulated to estimate $\mathbf{x}$ as well as the mapping kernel $f(.)$ relating $\mathbf{x}$ and $\mathbf{y}$, (ii) Acquisition physics based signal scaling is proposed to synthesize the desired image contrast similar to HF image, (iii) the simulated contrast image is used as a regularizer while estimating $\mathbf{x}$ from $\mathbf{y}$. 
The experimental results demonstrate that the proposed approach provides improved quality of MRI images in terms of image contrast and sharp tissue boundaries that is similar to HF images. The experiments further demonstrate the successful application of the improved quality images provided by proposed method in improved tissue segmentation and volume quantification.

\section{Proposed Method}
The proposed work formulates the estimation of HF image ($\mathbf{x}$) from LF image ($\mathbf{y}$) in an alternate minimization framework: 
\begin{equation}
    \begin{split}
    {\min_{\mathbf{h,x}}\underbrace{||\mathbf{y}-\mathbf{h}\ast\mathbf{x}||_F^2}_\text{Data Fidelity}, \   +\lambda_1\underbrace{ ||\mathbf{c}\odot \mathbf{y}-\mathbf{x}||_F^2}_\text{Physics based Regularizer}    + \lambda_2 \underbrace{||\mathbf{h}||^2_{2}}_\text{Regularizing $\mathbf{h}$}}  
    \end{split}
    \label{eq:completeeq}
\end{equation} 
Here $\mathbf{x}\in\mathbb{R}^{m\times n}$  and $\mathbf{y}\in\mathbb{R}^{m\times n}$ represent the HF and LF MRI images, respectively. The matrix $\mathbf{h}\in\mathbb{R}^{p\times p}$ represents  transformation kernel convolved using $\ast$ operator with each patch of $\mathbf{y}$ of size $p \times p$ and $p$ is empirically chosen as 5. The matrix $\mathbf{c}\in\mathbb{R}^{m \times n}$ represent the pixel wise scale when multiplied with $\mathbf{y}$  generates the image with contrast similar to HF image. Here, $\mathbf{c} \odot \mathbf{y}$ represents the Hadamard product of pixel wise scale $\mathbf{c}$ and $\mathbf{y}$ the 1.5T (or 0.25T) image. 

\subsection{Physics based Regularizer}
The physics based regularizer exploits the fact that 
the T1 relaxation time increases with FS. The differences among T1 relaxation times of gray matter (GM), white matter (WM) and cereberospinal fluid (CSF) also increase with FS, leading to increased contrast in the T1-weighted images in the order 3T $\geq$1.5T $\geq$ 0.25T. This factor is used to simulate an image from $\mathbf{y}$ which convey similar information as $\mathbf{x}$, and is obtained by scaling the signal intensities of $\mathbf{y}$ based on changes in signal due to changes in T1 relaxation time with respect to FS- denoted by $\mathbf{r}$. The 
acquired signal for $\mathbf{y}$ denoted by  $\mathbf{s}_l$ can be corrected with relaxation time to simulate the signal $\mathbf{s}_h$ that would have been acquired for $\mathbf{x}$ as:  
\begin{equation}
    \hat{\mathbf{s}}_h=\mathbf{r}\times \mathbf{s}_l.
    \label{eq:PBC}
\end{equation}
For example, in the spin echo (SE) pulse sequence, acquired signal  can be represented as, $\mathbf{s}=\mathbf{A}{B}_0^2 \frac{\left(1- {e} ^ {(-TR/T_1)} sin {\theta}\right)}{\left (1- {e} ^ {(-TR/T_1)}  cos {\theta}\right )}   {e} ^ {(-TE/T_2)}$~\cite{minimumFSplosOneSignalScaling,TRTE_SE}. 
Here, $\mathbf{A}$ represents the proportionality constant. The field strengths for 3T (or 1.5T) and 1.5T (or 0.25T) scanners are denoted by ${B}_{0}$. The factor $\mathbf{r}$ for the given voxel for SE sequence by assuming long TE, and T2 to be same across different FS (assumption derived from literature~\cite{goldentableT1}) be computed as 
\begin{equation}
{\frac{\left(1- {e} ^ {(-TR_h/T_{1,h})} sin {\theta_h}\right)}{\left (1- {e} ^ {(-TR_h/T_{1,h})}  cos {\theta_h}\right )}}/{\frac{\left(1- {e} ^ {(-TR_l/T_{1,l})} sin {\theta_l}\right)}{\left (1- {e} ^ {(-TR_l/T_{1,l})}  cos {\theta_l}\right )}}.
\label{eq:computer}
\end{equation}
Here, $T_1$ and $T_2$  represent the T1/T2 relaxation times, $TR$-repetition time, $TE$-echo time and $\theta$-flip angle (FA). The parameters with subscript $\textit{h}$ and $\textit{l}$ represent the parameters used for HF and LF MRI acquisitions, respectively. Similarly, respective mathematical formulations can be used for other pulse sequences such as fast SE (FSE) and gradient echo (GRE). After computing $\mathbf{r}$ using eq.(\ref{eq:computer}), it can be used to simulate HF acquisition $\mathbf{\hat{s}}_h$ using eq.(2)

The values of $T_1$/$T_2$ relaxation times differ with tissues, hence different values of $\mathbf{r}$ should be computed for every voxel considering the tissues present in that voxel. However, if we assign single tissue per voxel using any tissue segmentation technique such as FAST~\cite{FSL}, compute $\mathbf{r}$ for each voxel to estimate $\mathbf{\hat{s}}_h$, this could lead to discontinuity among tissue boundaries in simulated HF image, 
and is shown in Fig.~\ref{fig:3tissuesdiscontinuity} in supplementary material (SM).  

\subsubsection{Compute relaxation times for voxels with more than one tissue:} 
\begin{figure}
    \centering
    \includegraphics[width=\linewidth]{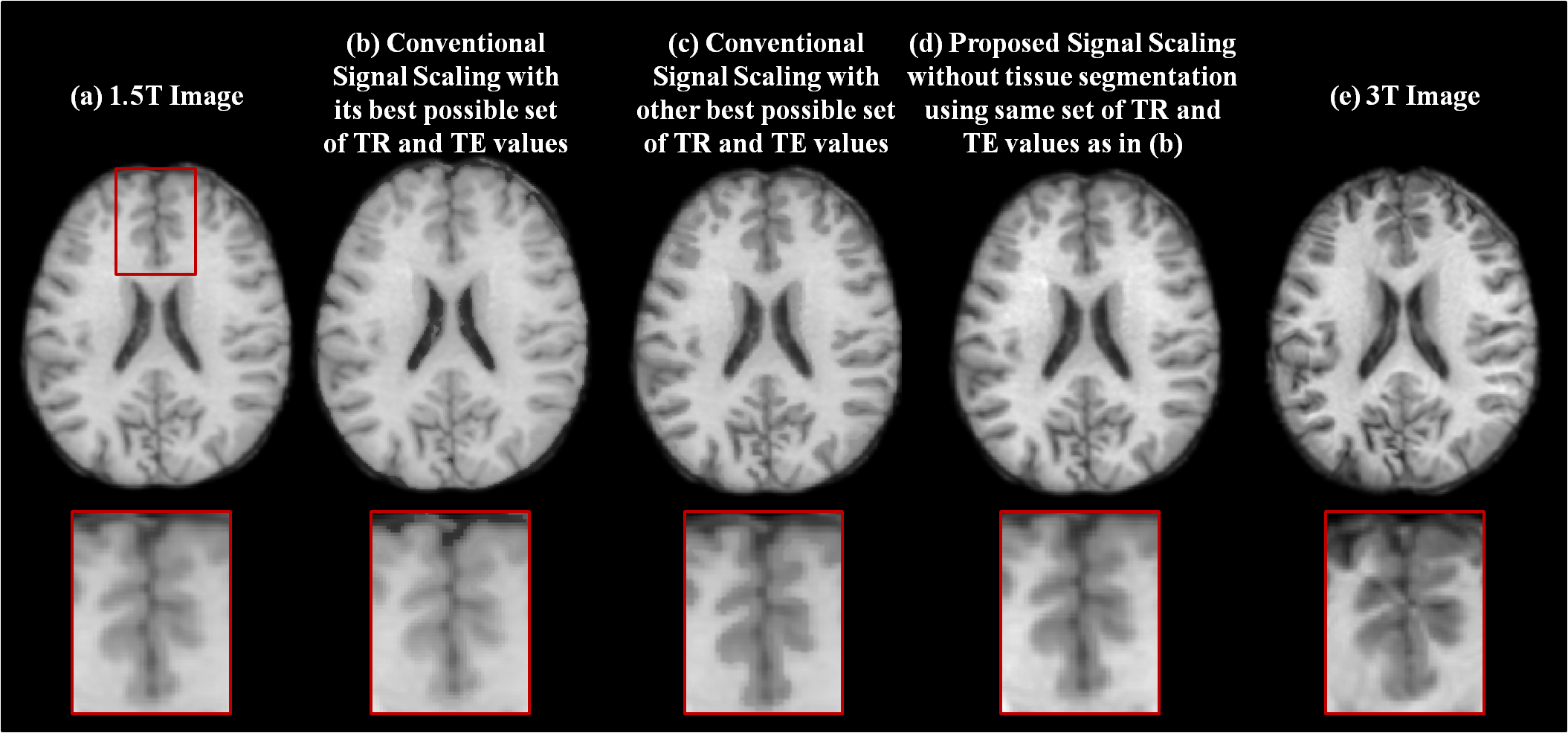}
    \caption{Demonstration of conventional signal scaling of 1.5T image shown in (a). Conventional signal scaling: Consider single tissue per voxel and tissues can be either of WM, GM, CSF that are scaled by computing $r$ and are shown in (b) and (c). (d) Proposed signal scaling approach by considering the voxels with more than one tissue, and computing $r$ for each of voxels with different proportions of tissues present, (e) Corresponding 3T MR image}
    \label{fig:3tissuesdiscontinuity}
\end{figure}

In real practice there exist many voxels with more than one tissue kind present in them, and is the reason for disconitinuities present in Fig.~\ref{fig:3tissuesdiscontinuity}. The challenge is that T1 relaxation time for such voxels is not known that is required to estimate $\mathbf{r}$. 
We address this issue by estimating the T1 relaxation time of such voxel as linear combination of T1 relaxation times of tissues present in the given voxel. The linear weights are directly proportional to the probability of tissues present in the voxel. 
Though this work performs well with probability maps but we use pixel intensity to denote the probability of tissue to avoid the additional and expensive tissue segmentation step as follows: The top 5 percentile of pixel intensities are assumed to belong to WM and bottom 20 percentile as GM. Consider $w$ and ${q}$ as T1 relaxation time and corresponding pixel intensity, respectively. We here approximate relation between $w$ and $q$ using linear equation as $\frac{{w}_{WM}-{w}_{GM}}{{q}_{WM}-{q}_{GM}}=\frac{{w}-{w}_{GM}}{{q}-{q}_{GM}}$. Here, subscript indicates the tissue type. 
Here, $w_{WM}$ and $w_{GM}$ are T1 relaxation times which are currently assumed to be constant, and taken from literature~\cite{goldentableT1} 
whereas $q_{WM}$ and $q_{GM}$ are the pixel intensity values computed by averaging pixel intensities falling in mentioned percentiles for WM and GM, respectively. In a simplest case, say for an image $q_{WM}$ and $q_{GM}$ are 1 and 0, then the linear relationship is reduced to $w=(w_{WM}-w_{GM})q + w_{GM}$. Hence, for the given pixel intensity we can estimate the corresponding T1 relaxation time as a linear combination of T1 relaxation times of WM and GM. 
The range of pixel intensities in $\mathbf{y}$ is partitioned into several bins. 
This is followed by approximation of T1 relaxation times ($T_{1,l}$) for different bins for $\mathbf{y}$ by assuming $w_{WM}$  and $w_{GM}$ as T1 relaxation times at LF FS. In the similar way, $T1_{1,h}$ can be approximated.

After approximating $T_{1,l}$ and $T_{1,h}$ for all voxels (and empirically choosing the $TR_h$ and $\theta_h$) the $\mathbf{r}$ is computed using eq.(\ref{eq:computer}) and is used to estimate the HF acquisition using eq.(\ref{eq:PBC}). The estimated $\hat{\mathbf{s}}_h$ is used to compute the pixel wise scale $\mathbf{c}$ as  $\mathbf{c}=\hat{\mathbf{s}}_h/\mathbf{y}$, i.e., an element wise division operator. The computed $\mathbf{c}$ is used in eq.(\ref{eq:completeeq}) to constrain the solution space of estimated HF image $\mathbf{x}$.
\vspace{-0.5cm}
\section{Experimental Results}
The proposed work has been demonstrated for (i) estimating 3T-\textit{like} from 1.5T images, (ii) estimating 1.5T-\textit{like} from 0.25T images, (iii) evaluating accuracy of tissue segmentation using improved images in (i) and (ii), and (iv) comparing (i), (ii) and (iii) with existing methods. The results related to (ii) are summarized in SM due to space constraints. The values for $\lambda_1$ and $\lambda_2$ are chosen as 1.2 and 0.4, respectively.
\subsection{Data}
The MRI images used to demonstrate the efficacy of proposed work were acquired from five healthy subjects of age $25\pm10$ years. Three different MRI scanners were used in this study: 0.25T (G-scan Brio, Esaote), 1.5T (Aera Siemens) and 3T (Verio, Siemens). The first three subjects were scanned using each of the three scanners while the other two subjects were scanned with only 1.5T and 3T scanners. The three scanners were located in Post Graduate Institute of Medical Education \& Research (PGIMER) Chandigarh, India. Scanning was performed using the standard clinical protocols-optimized for both clinical requirement and work-load of clinical site. 
All the scans were performed according to the guidelines of the Declaration of Helsinki. The details of pulse sequence and scan parameters used to acquire data in each of the three scanners is mentioned in SM Table~\ref{tab:acq}. The acquired T1 
MR image volumes for each scanner and each human subject were pre-processed similar to the human connectome project (HCP) pre-processing pipelines for structural MR images~\cite{HCPpreproc}. Please note that proposed approach does not require the LF and HF images to be skull stripped or to have pixel to pixel correspondence. It is only done to provide reference based similarity scores of estimated image with respect to HF image. 

\begin{figure}
\vspace{-0.5cm}
    \centering
    \includegraphics[width=0.75\linewidth]{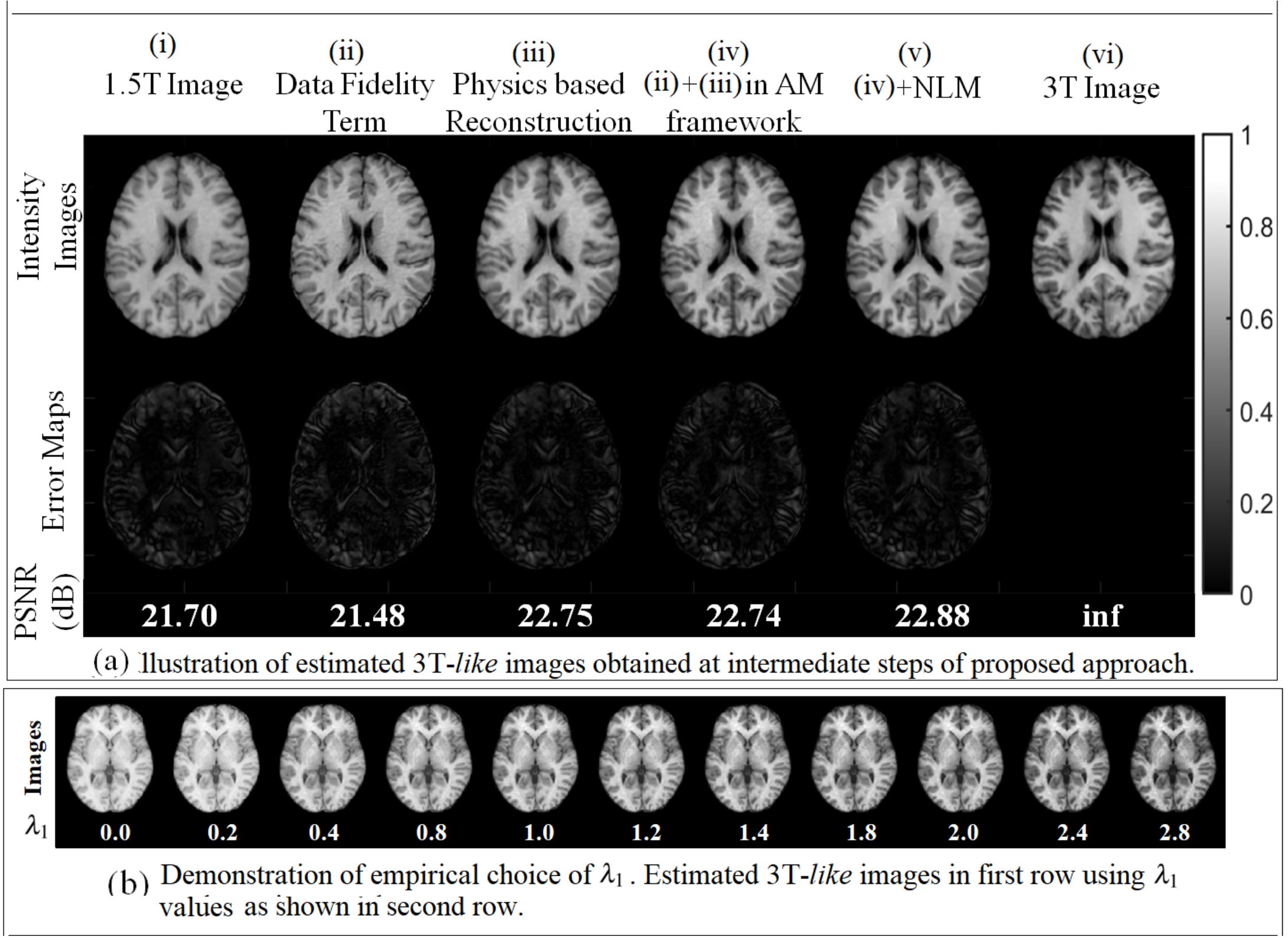}
    \caption{(a) Ablation study of proposed approach. (b) Demonstration of significance of proposed physics based constraint and associated parameter $\lambda_1$. }
    \label{fig:ablation}
    \vspace{-1cm}
\end{figure}

\subsection{Analysis/Ablation study of proposed approach}
The HF image is estimated at different stages of proposed approach to demonstrate the significance of each term in eq.(\ref{eq:completeeq}), and is shown in Fig.~\ref{fig:ablation} (a). In Fig.~\ref{fig:ablation} (b), the impact of proposed regularizer on estimation of HF image is demonstrated by changing the values of $\lambda_1$.  
It can be observed that the image Fig.~\ref{fig:ablation} (a)-(ii) obtained just from the data fidelity term leads to sharp image details but without any contrast improvement. However, HF image simulated by the physics based regularizer from eq.(\ref{eq:PBC}) improves the contrast but details remain blurred, as is evident from image Fig.~\ref{fig:ablation} (a)-(iii). Once the data fidelity and the physics based regularizer is combined as in eq.(\ref{eq:completeeq}) in AM framework the corresponding image is shown in Fig.~\ref{fig:ablation} (a)-(iv) that is sharper as well as improved in contrast, and with image sharpness=3.19, PIQE=4.7, and SSIM=0.0762. The image in Fig.~\ref{fig:ablation} (a)-(iv) is further smoothed in Fig.~\ref{fig:ablation} (a)-(v) using Non local means (NLM) approach to avoid any grainy effect if present due to division of pixel intensities into bins. The improvement in image quality is also evident from the PSNR values which increase from Fig.~\ref{fig:ablation} (a)-(ii) to (v). 
\begin{figure}
\vspace{-0.5cm}
    \centering
    \includegraphics[width=0.8\linewidth]{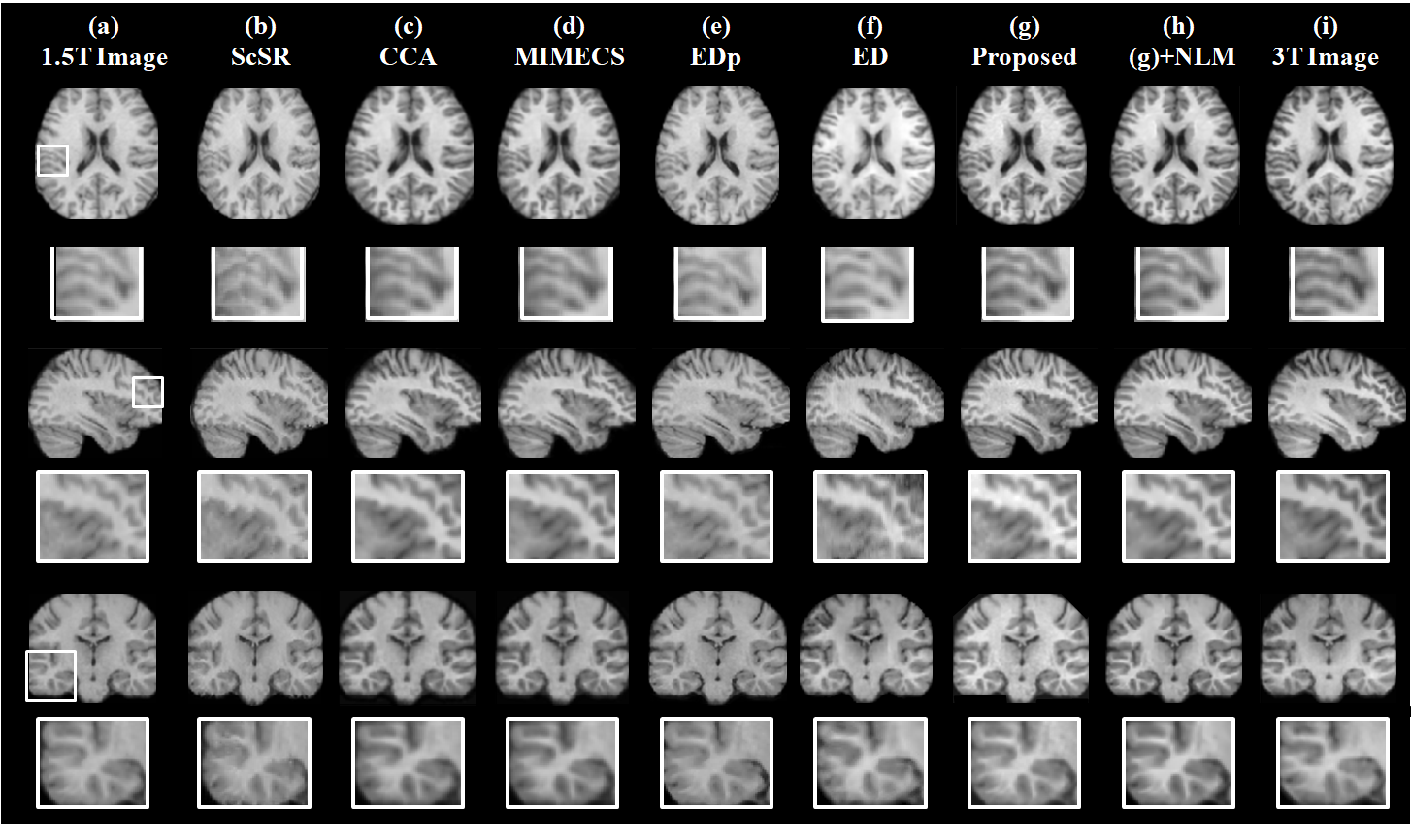}
    \caption{Comparison of quality of HF images estimated by various methods.}
    \label{fig:comaprison}
    \vspace{-1.0cm}
\end{figure}
\subsection{Comparison with Existing Approaches}
\begin{figure}
    \centering
    \includegraphics[width=\linewidth]
    {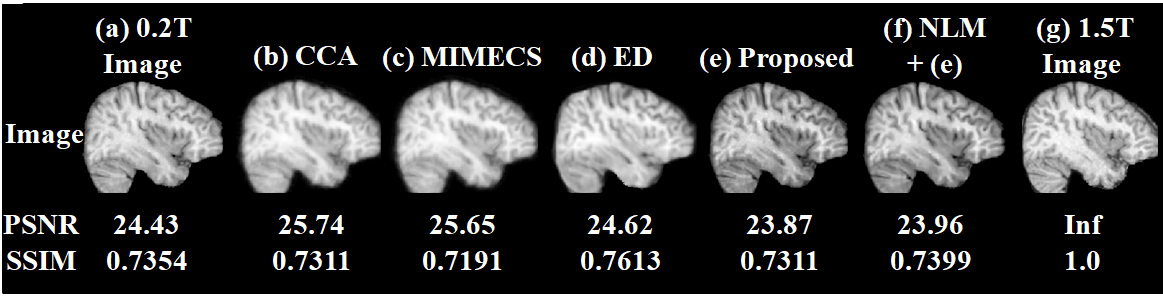}
    \caption{Comparison in visual appearance of 1.5T-like image details obtained using proposed method with existing approaches}
    \label{fig:2TSOA}
\end{figure}

The performance of proposed approach is compared with existing methods that either address the contrast synthesis or estimation of HF images ScSR \cite{YangWright},CCA \cite{cca}, MIMECS \cite{mimecs}, ED \cite{ICIP18}. The comparison is done in four ways: (i) {\textit{Objective} analysis using \textit{reference based} metrics }that requires the ground truth HF image, and pixel to pixel to correspondence between query image and ground truth image- Table-I, (ii) {\textit{Objective} analysis using \textit{no-reference based} metrics} which describe the quality of images solely based on edge sharpness and image contrast-  Table-II,(iii) \textit{Subjective analysis} that includes rating of images by clinical experts in range 0 to 5, 5 and 0 being the highest and worst quality images, respectively -  Table-III and (iv) \textit{Qualitative analysis } that includes the analysis of visual appearance of image details- Fig.~\ref{fig:comaprison}. It can be observed that the existing methods provide better performance in terms of reference based metrics but perform inferior to proposed method in case of no-reference based metrics and subjective scores. This is due to the way the existing methods are designed, i.e., these methods are trained to minimize the mean square error between estimated and HF images, thus they provide higher peak signal to noise ratio (PSNR) and structural similarity index metric (SSIM) but the image details are still blurred which lead to drop in edge sharpness. The validation of argument can also be derived by visually inspecting the images in Fig~\ref{fig:comaprison}. It has been observed that encoder-decoder based approach (ED~\cite{ICIP18}), MIMECS~\cite{mimecs}, CCA~\cite{cca} and ScSR~\cite{YangWright} provides blurred image details. The possible reasons are (i) minimizing MSE canprovide perceptually blurred results, (ii) the weighted averaging involved in \cite{mimecs,cca,YangWright} induces blur, (iii) inaccurate pixel to pixel correspondences makes it difficult for supervised methods to learn the actual mapping relating input and target images. The drop in performances of existing approaches due to inaccuracies in image registration is more prominently observed when improving 0.25T images in Fig.~\ref{fig:2TSOA}. The robustness to such inaccuracies by proposed approach due to its unsupervised nature shows its clear advantages over existing methods. 


\begin{figure}
    \vspace{-0.75cm}
    \centering
    \includegraphics[width=0.75\linewidth]{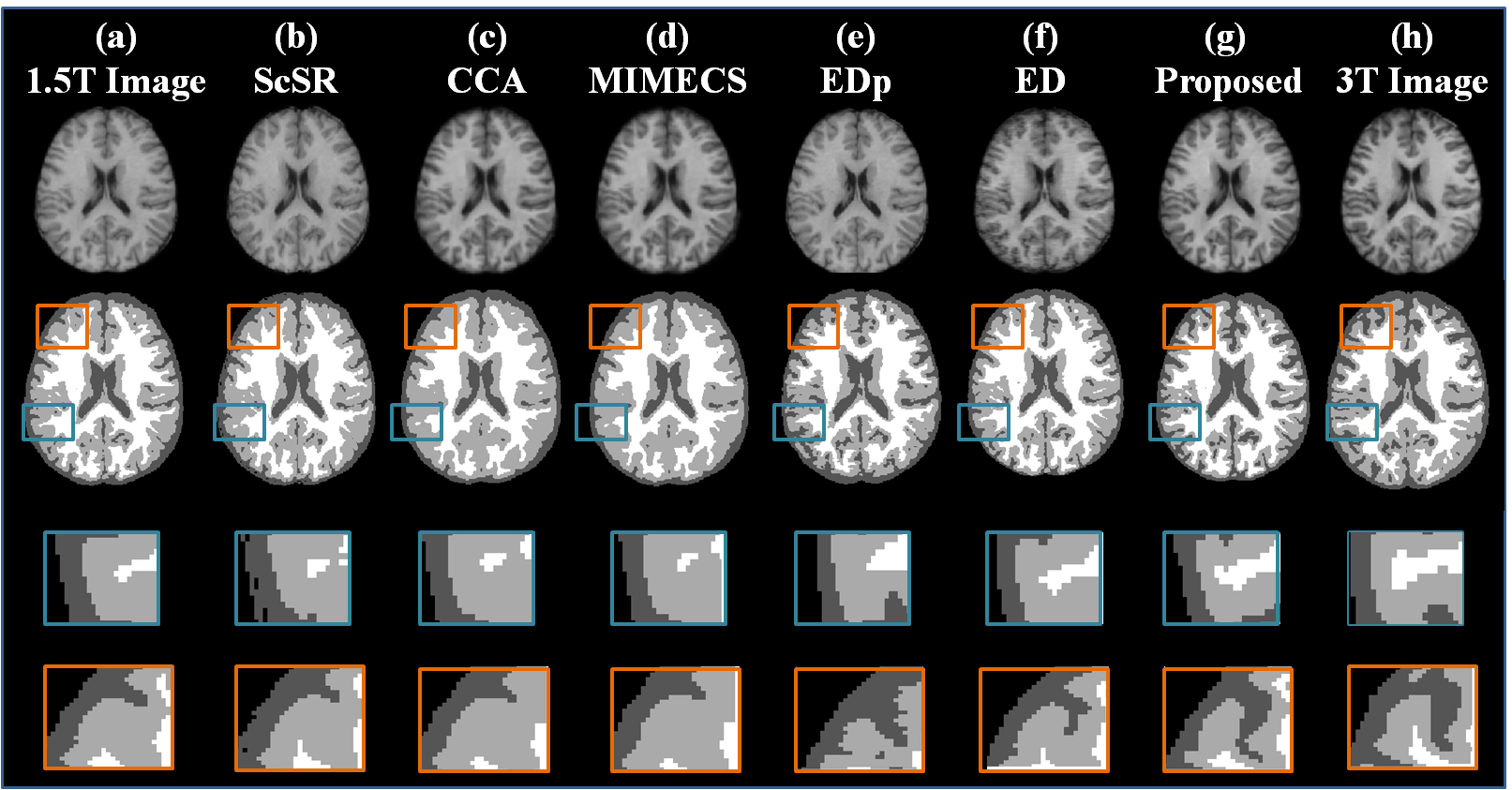}
    \caption{Demonstration of improved segmentation from images estimated by proposed approach and its comparison with existing approaches}
    \label{fig:segmentation}
    \vspace{-1.2cm}
\end{figure}

\subsection{Application to tissue segmentation and volume quantification}
\begin{figure}
    \centering
    \includegraphics[width=\linewidth]{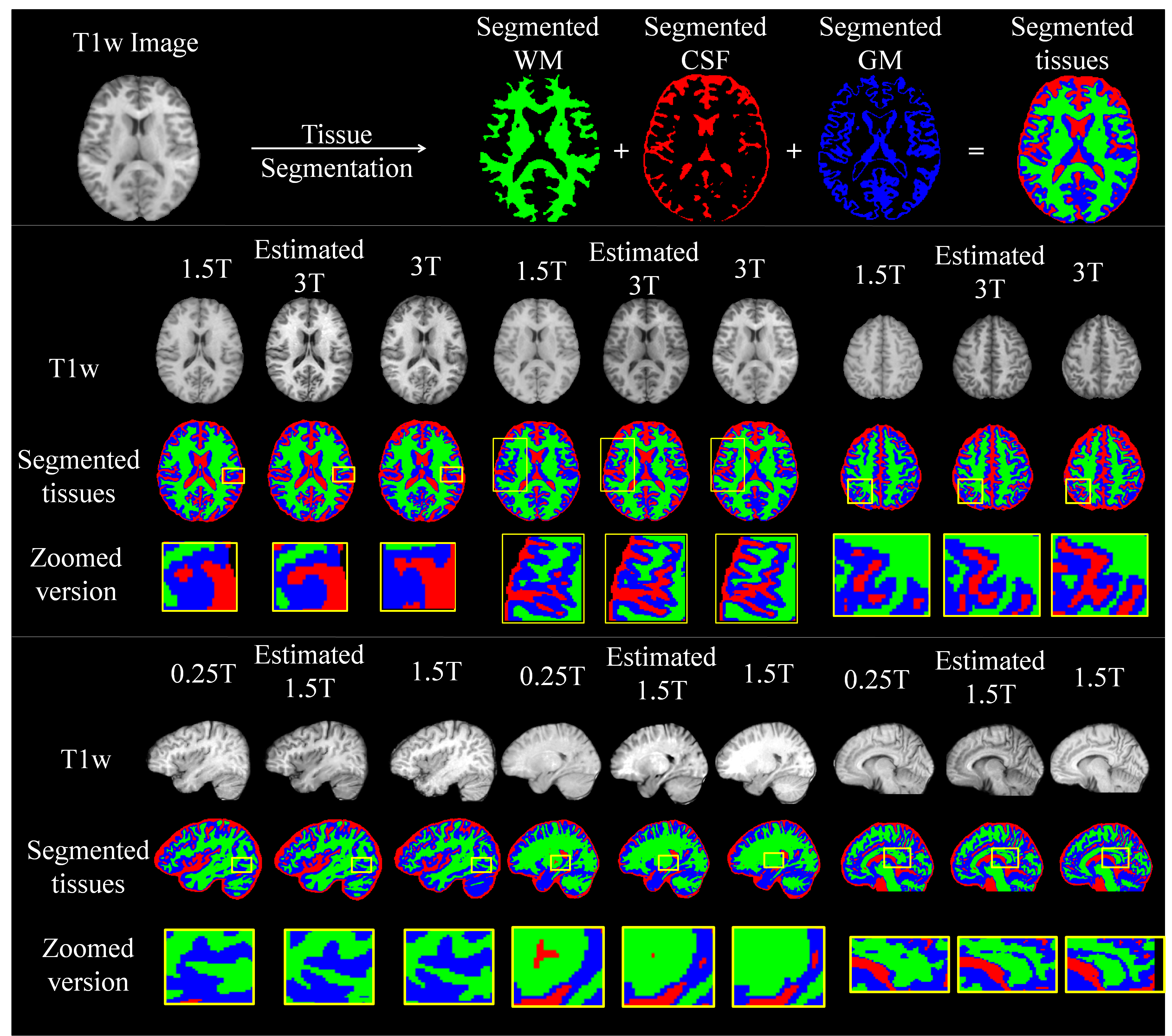}
    \caption{ Demonstration of improved tissue (GM, WM and CSF) segmentation using 3T-like (or 1.5T-like) MR images estimated using proposed method from 1.5T (or 0.25T) MRI of different locations}
    \label{fig:TissueSeg15T02T}
\end{figure}

\begin{figure}
    \centering
    \includegraphics[width=\linewidth]{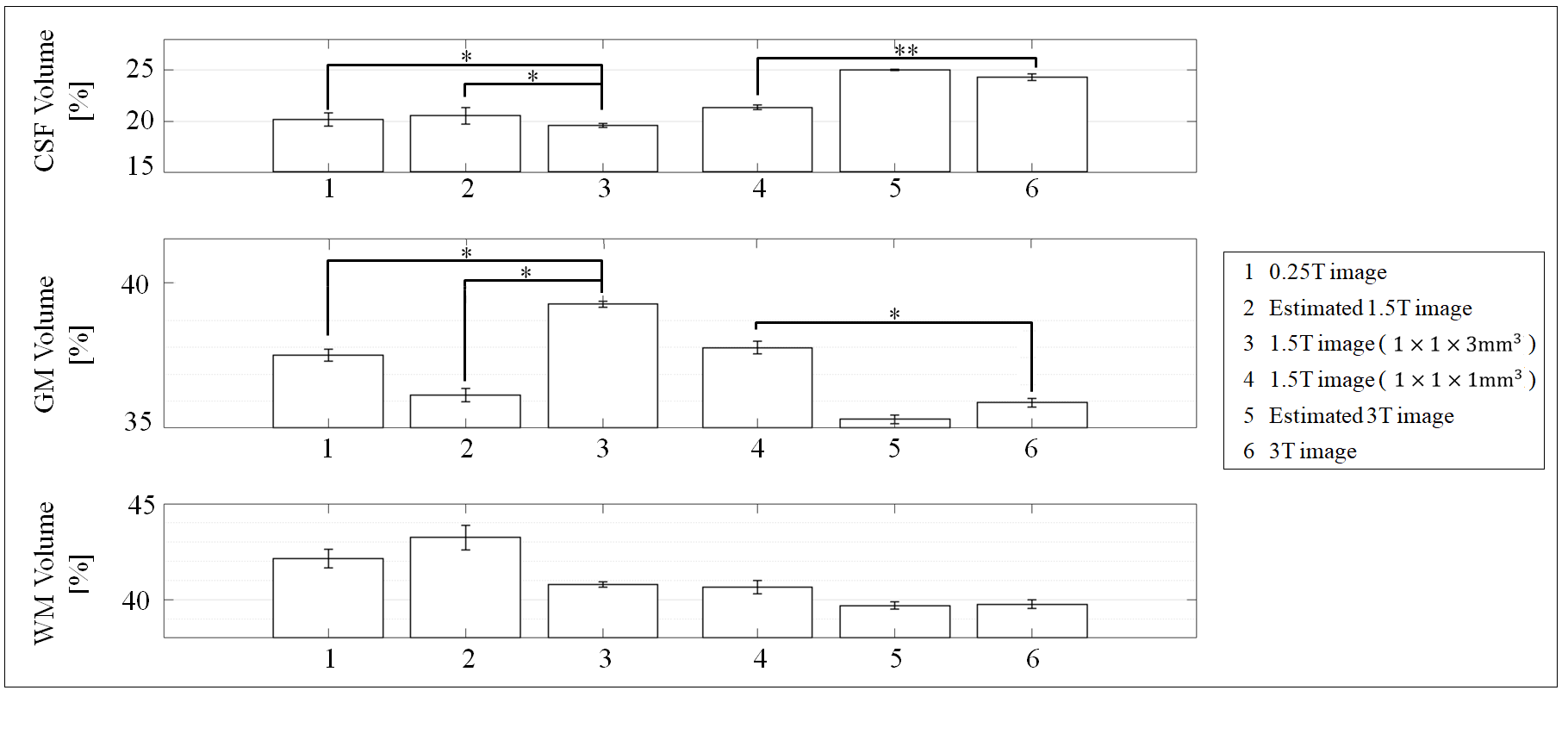}
    \caption{Barplots with error bars displaying the quantified tissue volumes for WM, GM and CSF from scanner acquired 1.5T (or 0.25T), estimated 3T-like (or 1.5T-like) and scanner acquired 3T (or 1.5T) images. Here, * denotes the validation of hypothesis H0:two input distributions are similar with 0.95 confidence level.}
    \label{fig:TissueSegQuant}
\end{figure}

The segmentation labels for WM, GM and CSF were computed using FAST toolbox in FSL software~\cite{FSL} for 3T-\textit{like} images estimated by different approaches, and shown in Fig.~\ref{fig:segmentation}. The improved tissue segmentation for 0.25T images is shown in Fig.~\ref{fig:TissueSeg15T02T} in SM. The zoomed windows in both figures indicate that the segmentation label of WM is improved for the estimated 3T reconstructed image by the proposed approach for estimated 3T image from 1.5T image. 
The quantitative measure used to evaluate performance of different methods is dice ratio, and is reported in Table~\ref{tab:diceratio}, and its comparison is mentioned in Table~\ref{tab:dicecomapre}. 
The ability to accurately segment tissues from image estimated by proposed approach is shown to be comparable both qualitatively and quantitatively to existing methods. 
Further, proposed method is shown to provide statistically significant with p$<$0.01 improvement in accuracy of WM and GM tissue volume quantification for estimated 3T (and 1.5T images), and is shown in Fig.~\ref{fig:TissueSegQuant}. 

\section{Summary}
We propose a method to estimate HF images from $\leq$1.5T images in an unsupervised manner. Here, the knowledge of acquisition physics to simulate HF image is exploited, and used it in a novel way to regularize the estimation of HF image. The proposed method demonstrates the benefits over state of the art supervised methods that are severely effected by the inaccuracies if present in image registration process. Lower the FS image is, harder is to get accurate image registration, and thus proposed method proves to be a better choice. 
Further, it is also demonstrated that the proposed approach provides statistically significant accurate tissue segmentation. The code for this work is publicly shared 
\href{https://drive.google.com/drive/folders/1WbzkBJS1BWAje8aF0ty2SWYTQ9i0B7Yr?usp=sharing} here.\\ 
\begin{adjustbox}{angle=90}
\resizebox{1.15\textwidth}{!}{
\begin{tabular}{|c|c|c|c|c|c|c|c|c|c|c|c|}
\multicolumn{12}{c}{\textbf{TABLE I}} \\ 
\multicolumn{12}{c}{\textbf{OBJECTIVE ANALYSIS- NO-REFERENCE BASED METRICS}} \\        

\hline
Metric            &             & 1.5T   & HM     & \multicolumn{5}{|c|}{Supervised Approaches}  &     \multicolumn{2}{|c|}{Unsupervised Approaches}               & 3T     \\ \hline
                  &             &        &        & SCSR & CCA  & MIMECS & EDp & ED & Proposed                & Proposed (NLM) &        \\ \hline
Signal Difference & WM-GM       & 0.1652 & 0.3414 & 0.1605                 & 0.2079                & 0.2084                   & 0.1291                & 0.1715               & 0.2166                  & 0.2165         & 0.2449 \\ \hline
                  & GM-CSF      & 0.3381 & 0.4536 & 0.3263                 & 0.3154                & 0.3112                   & 0.3345                & 0.3208               & 0.3207                  & 0.3209         & 0.3144 \\ \hline
Image Sharpness   & Sharpness   & 55.62  & 95.16  & 58.09                  & 54.81                 & 54.25                    & 60.39                 & 52.06                & 63.01                   & 64.13          & 64.6   \\ \hline
                  & Edge width  & 0.1634 & 0.1559 & 0.1527                 & 0.1837                & 0.1855                   & 0.1545                & 0.1716               & 0.1812                  & 0.1735         & 0.1776 \\ \hline
                  & Edge height & 9.08   & 14.81  & 8.87                   & 10.07                 & 10.06                    & 9.32                  & 8.93                 & 11.42                   & 11.12          & 11.46  \\ \hline
PIQE              & mean        & 67.57  & 60.63  & 51.52                  & 74.32                 & 75.86                    & 58.76                 & 82.81                & 57.78                   & 72.53          & 70.25  \\ \hline
\multicolumn{12}{c}{\textbf{}}       \\
\multicolumn{12}{c}{\textbf{TABLE II}}       
\\
\multicolumn{12}{c}{\textbf{OBJECTIVE ANALYSIS- REFERENCE BASED METRICS}}       
\\ \hline
Metric            &             & 1.5T   & HM     & \multicolumn{5}{|c|}{Supervised Approaches}  &     \multicolumn{2}{|c|}{Unsupervised Approaches}               & 3T     \\ \hline
                  &             &        &        & SCSR & CCA  & MIMECS & EDp & ED & Proposed                & Proposed (NLM) &        \\ \hline
PSNR   & mean & 23.6   & 17.71  & 22.57                  & 25.01                 & 25.02                    & 19.96                 & 25.41                & 24.24                   & 24.27          & inf \\ \hline
       & std. & 0.1873 & 0.2244 & 0.1639                 & 0.1441                & 0.1639                   & 0.0426                & 0.1178               & 0.1849                  & 0.1864         & -   \\ \hline
SSIM   & mean & 0.8314 & 0.8885 & 0.7639                 & 0.8531                & 0.8465                   & 0.6917                & 0.8511               & 0.8387                  & 0.8465         & 1   \\ \hline
       & std. & 0.0069 & 0.002  & 0.0047                 & 0.0066                & 0.0072                   & 0.0036                & 0.0058               & 0.0064                  & 0.0064         & -   \\ \hline
UQI    & mean & 0.4854 & 0.8581 & 0..4300                & 0.6647                & 0.6607                   & 0.4362                & 0.7053               & 0.5196                  & 0.5263         & 1   \\ \hline
       & std. & 0.101  & 0.0044 & 0.084                  & 0.0275                & 0.0265                   & 0.0065                & 0.0183               & 0.0755                  & 0.095          & -   \\ \hline
VIF    & mean & 0.2697 & 0.5022 & 0.188                  & 0.281                 & 0.282                    & 0.2784                & 0.3072               & 0.2451                  & 0.251          & 1   \\ \hline
       & std. & 0.0052 & 0.0074 & 0.002                  & 0.0063                & 0.0063                   & 0.0011                & 0.0067               & 0.0039                  & 0.004          & -   \\ \hline
\multicolumn{12}{c}{\textbf{}}       
\\
\multicolumn{12}{c}{\textbf{ TABLE III}}       
\\

\multicolumn{12}{c}{\textbf{SUBJECTIVE ANALYSIS}}       
\\ \hline
\multicolumn{2}{|c|}{Expert}          & 1.5T   & HM     & \multicolumn{5}{|c|}{Supervised Approaches}  &     \multicolumn{2}{|c|}{Unsupervised Approaches}               & 3T     \\ \hline
              \multicolumn{2}{|c|}{}             &        &        & SCSR & CCA  & MIMECS & EDp & ED & Proposed                & Proposed (NLM) &        \\ \hline
 \multicolumn{2}{|c|}{(i)}    & 3.25 & 4    & 2.5                    & 4                     & 4                        & 2.25                  & 3                    & 4                       & 4.25           & 5   \\ \hline
 \multicolumn{2}{|c|}{(ii)}   & 1    & 3    & 3                      & 4                     & 3                        & 3                     & 2                    & 4                       & 4              & 5  \\ \hline
 \multicolumn{2}{|c|}{(iii)}  & 1.75 & 3.75 & 1.5                    & 0.75                  & 1.5                      & 0                     & 2.25                 & 5                       & 4              & 4   \\ \hline
 \multicolumn{2}{|c|}{(iv)}   & 3.5  & 2.75 & 2.5                    & 1.5                   & 2                        & 0                     & 2.25                 & 4.25                    & 3              & 4.25\\ \hline
 \multicolumn{2}{|c|}{(v)}    & 0.25 & 5    & 0.25                   & 2                     & 2.25                     & 3                     & 1.25                 & 5                       & 4.25           & 4.25\\ \hline

\end{tabular}
}
\end{adjustbox}

\bibliographystyle{splncs04}
\bibliography{refs}

\newpage
\beginsupplement

\begin{figure}
    \centering
    \includegraphics[width=\linewidth]{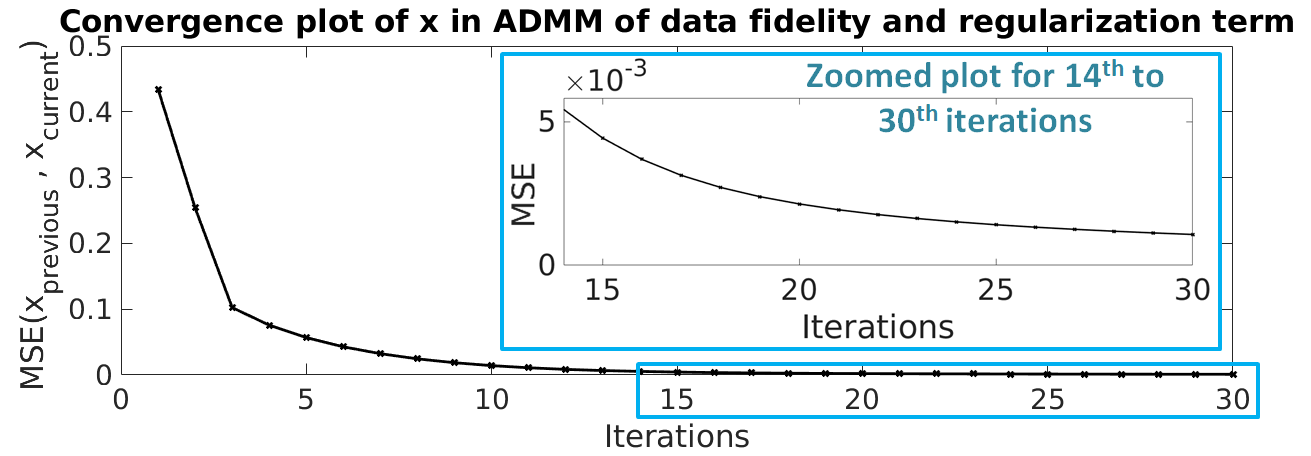}
    \caption{Convergence of optimization function in eq.(5) to estimate 3T-like MR image. Here, $\mathbf{x}_{previous}$  and $\mathbf{x}_{current}$ represent the estimated image $\mathbf{x}$ in previous and current iteration of eq.(5), respectively}
    \label{fig:my_label}
\end{figure}
\begin{figure}
    \centering
    \includegraphics[width=\linewidth]{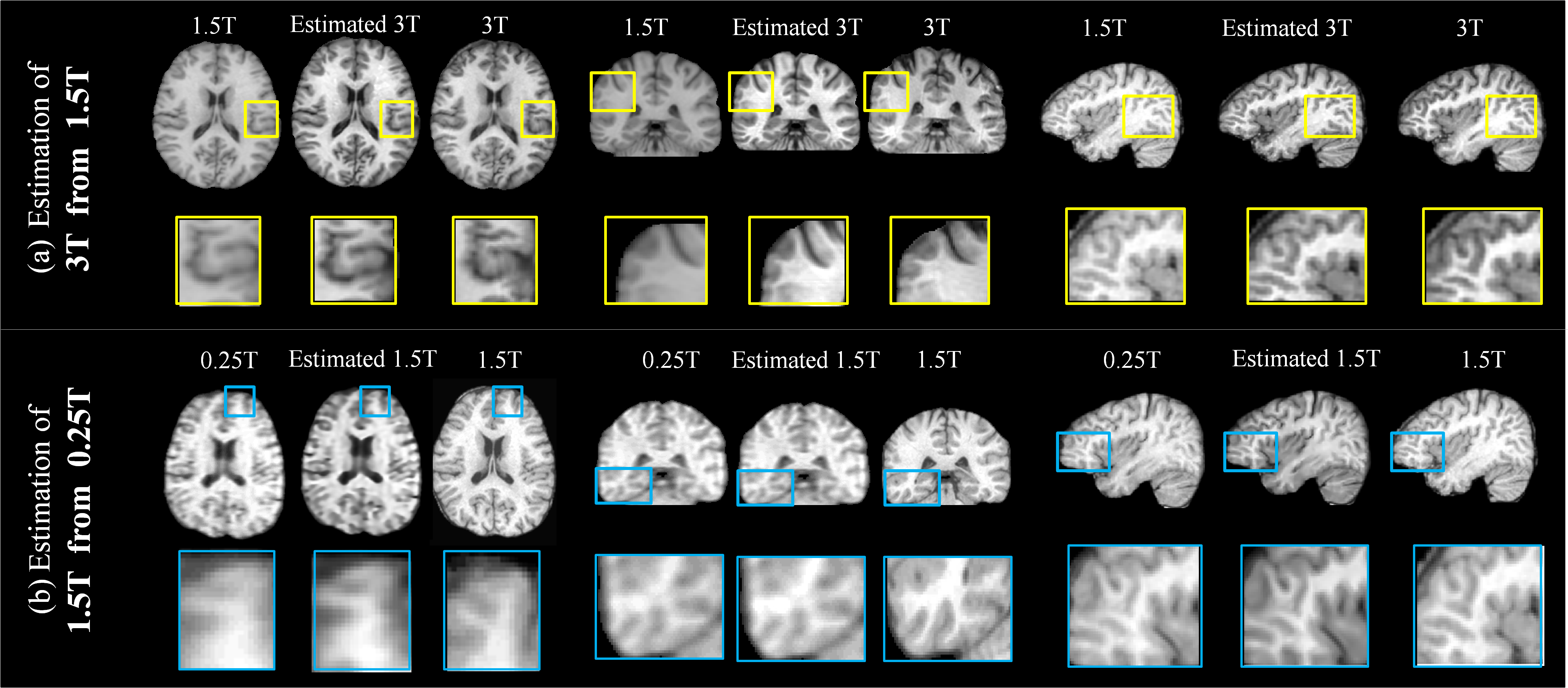}
    \caption{Demonstration of improved visual appearance of image details in 3T-like and 1.5T-like images estimated by proposed approach in three planes from 1.5T and 0.25T images in (a) and (b), respectively}
    \label{fig:my_label2}
\end{figure}

\begin{table}[]
\centering
\caption{Values used to acquire the images}
\label{tab:acq}
\begin{tabular}{|c|p{3cm}|p{3cm}|p{3cm}|}
\hline
           & 0.25T                            & 1.5T                                              & 3T                                             \\ \hline
Sequence   & SST1, TR=12ms, TE=6ms, FA=15$^\circ$, & MPRAGE, TR=1890ms, TE=4.17ms, TI=110ms, FA=15$^\circ$ & GRE, TR=8.428ms, TE=3.2ms, TI=450ms, FA=12$^\circ$ \\ \hline

resolution & 0.45x0.45x5mm$^3$         & 0.9x0.9x0.9mm$^3$                       & 1.08x1.08x1.08mm$^3$                   \\ \hline
\end{tabular}
\end{table}

\begin{table}[]
\centering
\caption{Dice ratio for WM, GM and CSF of scanner-acquired 1.5T (and 0.25T) and estimated 3T-like (and 1.5T-like) images}

\begin{tabular}{|c|c|c|c|}
\hline
{Estimation of 3T-like images}    & Tissue & Input 1.5T image  & Estimated 3T-like image   \\ \cline{2-4} 
                                                 & CSF    & 0.7462            & 0.7589                    \\ \cline{2-4} 
                                                 & GM     & 0.7653            & 0.7696                    \\ \cline{2-4} 
                                                 & WM     & 0.8489            & 0.8614                    \\ \hline
{Estimation of 1.5T-like  images} & Tissue & Input 0.25T image & Estimated 1.5T-like image \\ \cline{2-4} 
                                                 & CSF    & 0.7154            & 0.721                     \\ \cline{2-4} 
                                                 & GM     & 0.7259            & 0.7291                    \\ \cline{2-4} 
                                                 & WM     & 0.773             & 0.7964                    \\ \hline
\end{tabular}
\label{tab:diceratio}
\end{table}

\begin{table}[]
\centering
\caption{Dice ratio for WM, GM AND CSF of 3T MR images estimated by different approaches}

\begin{tabular}{|c|c|c|c|c|c|c|}
\hline
Tissue & 1.5T   & ScSR   & CCA    & MIMECS & ED     & Proposed \\ \hline
CSF    & 0.7462 & 0.7146 & 0.7371 & 0.74   & 0.7481 & 0.7589   \\ \hline
GM     & 0.7653 & 0.7385 & 0.7867 & 0.7881 & 0.7712 & 0.7696   \\ \hline
WM     & 0.8489 & 0.839  & 0.8638 & 0.8631 & 0.8643 & 0.8614   \\ \hline
\end{tabular}
\label{tab:dicecomapre}
\end{table}

\end{document}